\documentclass[onecolumn]{IEEEtran}

\usepackage[mathscr]{eucal}
\usepackage[cmex10]{amsmath}
\usepackage{epsfig,epsf,psfrag}
\usepackage{amssymb,amsmath,amsthm,amsfonts,latexsym}
\usepackage{amsmath,graphicx,bm,xcolor,url}
\usepackage[caption=false]{subfig} 
\usepackage{fixltx2e}%ordering of single and double column floats
\usepackage{array}%array and tabular environments
\usepackage{verbatim}
\usepackage{bm}
\usepackage{algorithmic, cite}
\usepackage{algorithm}
\usepackage{verbatim}
\usepackage{textcomp}
\usepackage{mathrsfs}
\usepackage{multirow}
\usepackage{bbm}
\usepackage{epstopdf}
\catcode`~=11 \def\UrlSpecials{\do\~{\kern -.15em\lower .7ex\hbox{~}\kern .04em}} \catcode`~=13 

\allowdisplaybreaks[1]
 
\newcommand{\nn}{\nonumber}

\newcommand{\calA}{\mathcal{A}}
\newcommand{\calB}{\mathcal{B}}

\newcommand{\calH}{\mathcal{H}}

\newcommand{\calT}{\mathcal{T}}

\newcommand{\calV}{\mathcal{V}}

\newcommand{\rmA}{\mathrm{A}}

\newcommand{\rmd}{\mathrm{d}}

\newcommand{\rme}{\mathrm{e}}

\newcommand{\rmi}{\mathrm{i}}

\newcommand{\rmm}{\mathrm{m}}

\newcommand{\rmn}{\mathrm{n}}

\newcommand{\rmQ}{\mathrm{Q}}

\newcommand{\rmV}{\mathrm{V}}

\newcommand{\bbE}{\mathbb{E}}

\newcommand{\bbN}{\mathbb{N}}

\newcommand{\bbP}{\mathbb{P}}

\newcommand{\bbR}{\mathbb{R}}

\DeclareMathAlphabet{\mathbsf}{OT1}{cmss}{bx}{n}
\DeclareMathAlphabet{\mathssf}{OT1}{cmss}{m}{sl}% slanted sans serif

\newcommand{\rvP}{\mathsf{P}}

\newcommand{\rvR}{\mathsf{R}}

\DeclareSymbolFont{bsfletters}{OT1}{cmss}{bx}{n}  
\DeclareSymbolFont{ssfletters}{OT1}{cmss}{m}{n}
\DeclareMathSymbol{\bsfGamma}{0}{bsfletters}{'000}
\DeclareMathSymbol{\ssfGamma}{0}{ssfletters}{'000}
\DeclareMathSymbol{\bsfDelta}{0}{bsfletters}{'001}
\DeclareMathSymbol{\ssfDelta}{0}{ssfletters}{'001}
\DeclareMathSymbol{\bsfTheta}{0}{bsfletters}{'002}
\DeclareMathSymbol{\ssfTheta}{0}{ssfletters}{'002}
\DeclareMathSymbol{\bsfLambda}{0}{bsfletters}{'003}
\DeclareMathSymbol{\ssfLambda}{0}{ssfletters}{'003}
\DeclareMathSymbol{\bsfXi}{0}{bsfletters}{'004}
\DeclareMathSymbol{\ssfXi}{0}{ssfletters}{'004}
\DeclareMathSymbol{\bsfPi}{0}{bsfletters}{'005}
\DeclareMathSymbol{\ssfPi}{0}{ssfletters}{'005}
\DeclareMathSymbol{\bsfSigma}{0}{bsfletters}{'006}
\DeclareMathSymbol{\ssfSigma}{0}{ssfletters}{'006}
\DeclareMathSymbol{\bsfUpsilon}{0}{bsfletters}{'007}
\DeclareMathSymbol{\ssfUpsilon}{0}{ssfletters}{'007}
\DeclareMathSymbol{\bsfPhi}{0}{bsfletters}{'010}
\DeclareMathSymbol{\ssfPhi}{0}{ssfletters}{'010}
\DeclareMathSymbol{\bsfPsi}{0}{bsfletters}{'011}
\DeclareMathSymbol{\ssfPsi}{0}{ssfletters}{'011}
\DeclareMathSymbol{\bsfOmega}{0}{bsfletters}{'012}
\DeclareMathSymbol{\ssfOmega}{0}{ssfletters}{'012}

\newcommand{\tilA}{\tilde{A}}

\newcommand{\tilg}{\tilde{g}}

\newcommand{\hatv}{\hat{v}}
\newcommand{\hatV}{\hat{V}}
\newcommand{\tilv}{\tilde{v}}
\newcommand{\tilV}{\tilde{V}}

\newcommand{\barv}{\bar{v}}

\newcommand{\barV}{\bar{V}}

\newcommand{\eps}{\varepsilon}
\DeclareMathOperator{\var}{\mathrm{Var}}

\newcommand{\bone}{\mathbbm{1}}

\newtheorem{theorem}{Theorem} 
\newtheorem{lemma}{Lemma}

\newtheorem{definition}{Definition}

\newtheorem{remark}{Remark}

\newcommand{\qednew}{\nobreak \ifvmode \relax \else
      \ifdim\lastskip<1.5em \hskip-\lastskip
      \hskip1.5em plus0em minus0.5em \fi \nobreak
      \vrule height0.75em width0.5em depth0.25em\fi}

\usepackage{flushend}
\usepackage{dsfont}
\usepackage{xspace}
\usepackage[ colorlinks = true,
           linkcolor = blue,
            urlcolor  = blue,
            citecolor = red,
            anchorcolor = green,
]{hyperref}

\usepackage{cite}
\allowdisplaybreaks[2]
\begin{document}
\title{The Optimal Compression Rate of Variable-to-Fixed Length   Source Coding with a Non-Vanishing Excess-Distortion Probability} 

% %%% Single author, or several authors with same affiliation:
% \author{%
%   \IEEEauthorblockN{Stefan M.~Moser}
%   \IEEEauthorblockA{ETH Zürich\\
%                     ISI (D-ITET)\\
%                     CH-8092 Zürich, Switzerland\\
%                     Email: moser@isi.ee.ethz.ch}
% }

%%% Several authors with up to three affiliations:
\author{Lan V.\ Truong, {\em Member, IEEE}   $\quad$ 
        Vincent Y.~F.~Tan, {\em Senior Member, IEEE}  \thanks{The authors are with the  Department of Electrical and Computer Engineering, National University of Singapore (NUS). V.~Y.~F.~Tan is also with the Department of Mathematics, NUS. Emails: \url{lantruong@u.nus.edu};  \url{vtan@nus.edu.sg}} %\thanks{The authors are supported by an NUS Young Investigator Award (R-263-000-B37-133) and a  Singapore Ministry of Education (MOE) Tier 2 grant (R-263-000-B61-112).}
}

%%% Many authors with many affiliations:
% \author{%
%   \IEEEauthorblockN{Albus Dumbledore\IEEEauthorrefmark{1},
%                     Olympe Maxime\IEEEauthorrefmark{2},
%                     Stefan M.~Moser\IEEEauthorrefmark{3}\IEEEauthorrefmark{4},
%                     and Harry Potter\IEEEauthorrefmark{1}}
%   \IEEEauthorblockA{\IEEEauthorrefmark{1}%
%                     Hogwarts School of Witchcraft and Wizardry,
%                     1714 Hogsmeade, Scotland,
%                     \{dumbledore, potter\}@hogwarts.edu}
%   \IEEEauthorblockA{\IEEEauthorrefmark{2}%
%                     Beauxbatons Academy of Magic,
%                     1290 Pyrénées, France,
%                     maxime@beauxbatons.edu}
%   \IEEEauthorblockA{\IEEEauthorrefmark{3}%
%                     ETH Zürich, ISI (D-ITET), ETH Zentrum, 
%                     CH-8092 Zürich, Switzerland,
%                     moser@isi.ee.ethz.ch}
%   \IEEEauthorblockA{\IEEEauthorrefmark{4}%
%                     National Chiao Tung University (NCTU), 
%                     Hsinchu, Taiwan,
%                     moser@isi.ee.ethz.ch}
% }

\maketitle

%%%%%%
%% Abstract: 
%% If your paper is eligible for the student paper award, please add
%% the comment "THIS PAPER IS ELIGIBLE FOR THE STUDENT PAPER
%% AWARD." as a first line in the abstract. 
%% For the final version of the accepted paper, please do not forget
%% to remove this comment!
%%
\begin{abstract}
We consider the variable-to-fixed length lossy source coding (VFSC) problem. The optimal compression rate of the average length of variable-to-fixed source coding, allowing a non-vanishing probability of excess-distortion $\eps$, is shown to be equal to $(1-\eps)R(D)$, where $R(D)$ is the rate-distortion function of the source. In comparison to the related results of Koga and Yamamoto as well as Kostina, Polyanskiy, and Verd\'{u} for fixed-to-variable length  source coding, our results demonstrate an interesting feature that variable-to-fixed length source coding has the same first-order compression rate as fixed-to-variable length  source coding.
\end{abstract}
\section{Introduction}
 We investigate the fundamental limits of a variable-to-fixed length  lossy compression model for memoryless sources. The lossless variable-to-fixed length source coding without side information problem was first introduced by Tunstall~\cite{Tunstall} (also see the survey article~\cite{Abrahams}). In his Ph.D.\ thesis, Tunstall proposed a recursive,  entropy coding- and tree-based method  for lossless data compression that is also average-sense optimal. This means that this algorithm's compression rate is the entropy of the source.  Yamamoto and Yokoo~\cite{Yamamoto2001} proposed an algorithm to construct an average-sense optimal {\em almost instantaneous} variable-to-fixed length (AIVF) code for one-shot coding that attains the minimum average parse length for AIVF codes.  This class of codes has the advantage over regular Tunstall codes due to the fact that  almost instantaneous encoding is possible because the encoding delay is at most one symbol. Other authors~\cite{Sav97,Tjalkens87} extended the utility and analysis of Tunstall codes to Markov sources. A central limit theorem for the Tunstall algorithm's code length was derived in \cite{Drmota}. Universal versions of Tunstall's algorithm for various sources have also been suggested in~\cite{Tjalkens92,Vis01, Iri17}. In particular, Iri and Kosut~\cite{Iri17} derived the third-order coding rate for universal variable-to-fixed length lossless source coding for parametric sources belonging to an exponential family. 

For the dual problem---fixed-to-variable source coding---there are many known results, some of which we review here. For example, Koga and Yamamoto~\cite{Koga2005} characterized asymptotically achievable rates of variable-length prefix codes with non-vanishing error probability. The authors adopted an information spectrum approach~\cite{Han10}. Particularizing their general result to  finite alphabet i.i.d.\ sources with distribution $P_V$, they obtained 
\begin{align}
\label{eq1}
\lim_{N\to \infty} \frac{\ln M^*_{\mathrm{fv}}(N, \eps) }{N} =(1-\eps)H(V), \quad\forall\, \eps \in [0,1)
\end{align}
where $\ln M^*_{\mathrm{fv}}(N, \eps)$ is the minimum average length of a prefix-free code of the source $V^N$ subject to the reconstruction error probability being asymptotically at most $\eps$. By taking $\eps \to 0$,~\eqref{eq1} recovers a previous result by Han~\cite{Han2000a}. Recently, Kostina, Polyanskiy, and Verd\'{u}~\cite{kost17} quantified the minimal average code length attainable by lossy source-channel codes with variable-length feedback. The authors  concluded that such codes lead  to a significant improvement in the fundamental delay-distortion tradeoff. They showed that   separate source-channel coding fails to achieve these minimal average delays if a non-vanishing distortion probability is allowed. In particular, they showed that in fixed-to-variable length compression of a block of $N$ i.i.d.\ source outcomes,   under regularity assumptions,  the minimum average encoded length $\ln M^*_{\mathrm{fv}} (N, D, \eps)$ compatible with  the constraint that the probability of exceeding distortion $D$ is bounded by  $\eps$  satisfies
\begin{align}
\ln M^*_{\mathrm{fv}}(N,D,\eps)=(1-\eps)N R(D)-\sqrt{\frac{NV(D)}{2\pi}}\exp\bigg(-\frac{1}{2} \rmQ^{-1}(\eps)^2\bigg)+O(\ln N), \label{eqn:kpv}
\end{align}
where $R(D)$ and $V(D)$ are the rate-distortion and the rate-dispersion functions of $P_V$, and $\rmQ(t)=\int_t^\infty\frac{1}{\sqrt{2\pi}}\rme^{-u^2/2}\,\rmd u$ is the cumulative distribution function of the standard normal distribution. 

Our contribution in this paper is to extend the above studies in the following direction. We consider variable-to-fixed length codes for i.i.d.\ sources subject to some regularity conditions. Thus the class of codes we consider is similar to that considered by Tunstall~\cite{Tunstall}. However, in Tunstall's setting, no loss is allowed; the code must be designed so that it reconstructs the source perfectly. We allow for {\em lossy} reconstruction of the source. That is, we allow the probability of a certain separable distortion measure between the original and reconstructed source exceeding a threshold $D$ be no larger than a fixed constant $\eps\in (0,1]$. Under this more general scenario, we evaluate the {\em $\eps$-rate-distortion function}, i.e.,  the minimum rate of compression such that the source can be reconstructed up distortion level $D$ with probability of excess-distortion being upper bounded by $\eps$. We show that the $\eps$-rate-distortion function is $1-\eps$ times the usual rate-distortion function of the source $R(D)$. This first-order fundamental limit  is thus similar to the fixed-to-variable-length setting considered by Kostina, Polyanskiy, and Verd\'{u}~\cite{kost17} as can be seen from~\eqref{eqn:kpv}. We also establish bounds on the second-order terms. As a simple corollary of our result, it is shown that the Tunstall code~\cite{Tunstall} for variable-to-fixed length lossless source coding is (first-order) optimal (but this was already known from the results of~\cite{Drmota}).

\section{Problem Setting}\label{sec:dmc_bc}
Throughout, we let $\{V_n\}_{n=1}^{\infty}$ be an i.i.d.\ source with distribution $P_V$ and taking values in a set $\calV$ (finite or infinite).
\begin{definition} 
\label{def1} For any $N \in \bbR^+$, an $(M,N,D,\eps)$-{\em variable-to-fixed  length  source code (VFSC)} is defined by
\begin{itemize}
\item A sequence of source encoding functions $f_n: \calV^n\to \{1,2,\ldots,M\}$ where $n\in \bbN$.
\item A sequence of decoding functions $g_n: \{1,2,\ldots,M\} \to \hat{\calV}^n$ indexed by $ n\in \bbN$, each providing an estimate 
\begin{align}
\hatV^n =g_n(f_n(V^n)) 
\end{align}
at time $n$ at the decoder. Here,~$\hat{\calV}$ is the set of reproduction symbols.
\item An integer-valued random variable $\tau_N$ which is a stopping time of the filtration $\{\sigma(V^n)\}_{n=1}^{\infty}$ such that it satisfies
\begin{align}
\label{gh}
N \leq \bbE(\tau_N) \le N+o(N).
\end{align} 
\item The final decision (reconstruction) at the decoder $\hat{V}^{\tau_N} = (V_1,V_2, \ldots, V_{\tau_N})$ is computed at  the stopping time $\tau_N$ and $\hatV^{\tau_N}$ must satisfy the excess-distortion constraint
\begin{align}
\label{errordef}
\rvP_{\rmd}(N,D):=\bbP(d(\hat{V}^{\tau_N},V^{\tau_N})> D)\leq \eps,
\end{align}
where the separable distortion measure reads
\begin{align}
\label{measurepro1}
d(v^n,\hatv^n):=\frac{1}{n}\sum_{k=1}^n d(v_k,\hat{v}_k).
\end{align}
\end{itemize}
\end{definition}

We now comment on the somewhat unusual requirement on the stopping time $\tau_N$ in \eqref{gh}. For the VFSC problem, there is a tradeoff between the desire to compress as much as possible (this can be done by making  the stopping time  $\tau_N$ as large as possible) and desire to reduce the length of the source code (for this purpose, we make $\tau_N$ as small as possible). This leads to the constraint in~\eqref{gh}.  The formulation above, especially \eqref{gh}, is non-standard; however, it allows us to prove a tight result for the $\eps$-rate-distortion function for variable-to-fixed length lossy source coding. In particular, equipped with this formulation we can establish a converse for Tunstall coding~\cite{Tunstall}. 

We assume that the single-letter distortion measure $d: \calV \times \cal{\hatV} \to$  $[0,\infty)$  in~\eqref{measurepro1} satisfies
\begin{itemize}
\item (A1) We consider distortion  measures that are bounded below as follows:
\begin{align}
\label{maxdist}
D>d_{\rmm\rmi\rmn}:=\inf\{D>0: R(D)>0\}.
\end{align}
\item (A2)   Let the  {\em information density} and {\em mutual information} respectively be  defined as 
\begin{align}
\label{preinform}
\iota_{P_{V\hatV}}(v;\hatv)&:=\ln \frac{P_{\hatV|V}(\hatv|v)}{P_{\hatV}(\hatv)}, \enspace \forall (v,\hatv) \in \calV\times \hat{\calV},\quad \mbox{and}\\
I(V;\hatV)&:=\bbE_{P_{V\hatV}}\big[ \iota_{ P_{V\hatV}  }(V;\hatV) \big].
\end{align} 
 We assume that there exists a conditional distribution (test channel) $P_{\hatV|V}$ that achieves minimum in the {\em rate-distortion function}
\begin{align}
\label{distfunc}
R(D):=\min_{P_{\hatV|V} : \bbE_{P_{V\hatV}}[d( V,\hatV)]\leq D} I(V;\hatV)<\infty,
\end{align}
and \begin{align}
\label{eq10hold}
\rmV(D)&:=\var_{P_{V\hatV}}(\iota_{ P_{V\hatV}}(V;\hatV))<\infty,\\
\label{eq11hold}
\tilde{\rmV}(D)&:=\var_{P_{V\hatV}}(d(V,\hatV))<\infty.
\end{align}
\item (A3)  For $D>0$, $R(D)$ is continuously differentiable. Furthermore, $\rmV(D)$ and $\tilde{\rmV}(D)$  defined in \eqref{eq10hold} and \eqref{eq11hold} respectively are  assumed to be continuous.
\end{itemize}
Define the non-asymptotic fundamental limit 
\begin{align}
M_{\rm{vf}}^* (N,D,\eps):=\min\{M: \exists \mbox{ an $(M,N,D,\eps)$-VFSC}\}.
\end{align}
The asymptotic compression rate under a non-vanishing excess-distortion error probability of this variable-to-fixed  length source coding problem is defined as
\begin{align}
\label{codingrate}
R_{\rm{vf}}^*(D,\eps):=\limsup_{N\to \infty}\frac{\ln M_{\rm{vf}}^*(N,D,\eps) }{N}.
\end{align} 
Characterizing  $R_{\rm{vf}}^*(D,\eps)$, the {\em $\eps$-rate-distortion function} using variable-to-fixed length source codes,  is the central focus of this paper.
\section{Main Results}\label{sec:mainresult}
\begin{theorem} \label{thm} Assume that (A1)--(A3) hold. Then  for all $\eps\in (0,1]$,
\begin{align}
\label{eqmain}
R_{\rm{vf}}^*(D,\eps)= (1-\eps)R(D).
\end{align}
\end{theorem}
\begin{IEEEproof}
The achievability proof is in Section~\ref{ach:proof} and the converse proof is in Section~\ref{conv:proof}.
\end{IEEEproof}
 Some remarks are in order.
\begin{itemize}
\item  Theorem~\ref{thm} applies to both discrete and Gaussian memoryless sources. For the latter we naturally adopt the quadratic distortion measure  $d(v,\hatv)=(v-\hatv)^2$.
\item The assumptions (A1)--(A3)  are only used in  the achievability proof. The converse holds without assumption (A3).
\item In view of the results in~\cite{Koga2005},~\cite{Han2000}, and~\cite{KostinaVL}, Theorem~\ref{thm} suggests  that the performances of  variable-to-fixed and fixed-to-variable length  lossy source codes  are the same in the sense that the first-order compression rates are identical and equal to $(1-\eps)R(D)$. 
\item While Tunstall's algorithm~\cite{Tunstall} and its variants (e.g.,~\cite{Tjalkens92,Vis01}) are    constructive, we use random coding to upper bound  $R_{\rm{vf}}^*(D,\eps)$ by $(1-\eps)R(D)$. It remains to find a practical Tunstall-like algorithm that can realize this performance limit. However, Theorem \ref{thm} specialized to the case in which $\eps=D=0$  (so $R_{\rm{vf}}^*(D,\eps)=H(V)$) shows that the Tunstall code is optimal for variable-to-fixed length lossless source coding (cf.~\cite{Drmota}).
\item Our achievability and converse proofs  (see \eqref{goodness}  and \eqref{test2} to follow)    allow to obtain some bounds on the second-order term in the asymptotic expansion of $\ln M_{\rm{vf}}^*(N,D,\eps) $. However, these bounds are not tight, even in the order. 
%
% in fact imply that, for sufficiently large $N$, there exists positive constants $A_1,A_2>0$ such that 
%\begin{equation}
% -\frac{A_1}{\sqrt{N}} \le \frac{\ln M_{\rm{vf}}^*(N,D,\eps) }{N}-R(D)\le A_2\sqrt{ \frac{\ln N}{N}}.
%\end{equation}
We endeavored to establish the order of the second-order term but significantly new ideas in the proof technique and in the constraint on the stopping time in~\eqref{gh} are required to be successful in this direction.
\item In \cite{Drmota}, Drmota, Reznik and Szpankowski showed for the Tunstall code achieves the performance $\ln M_{\rm{vf}}^*(N,0,\eps) = N H(V)+ L_\eps\sqrt{N}  +o(\sqrt{N})$. The constant of proportionality $L_\eps$ was also identified.  They implicitly assumed that the stopping time $\tau_N$ satisfies $\bbE(\tau_N)=N+o(N)$ similarly to our formulation in~\eqref{gh}.% but the variance of the stopping time satisfies $\var(\tau_n)=\Theta(N)$. In our achievability proof, we fix the stopping time, hence $\var(\tau_n)=0$. Thus,  this is better than the Tunstall code. 

 \item In \cite{Iri17}, Iri and Kosut derived the second- and third-order asymptotics of universal variable-to-fixed length lossless source coding for parametric sources belonging to an exponential family. In contrast, we consider general memoryless sources  in this paper. We also note that the $\eps$-rate-distortion function defined  \eqref{codingrate} is different from the corresponding quantity defined in~\cite[Eq.~(7)]{Iri17}. In accordance to the setting in~\cite[Eq.~(7)]{Iri17}, Iri and Kosut showed in~\cite[Theorem 1]{Iri17} that the strong converse holds, but in our formulation,  the strong   converse does not hold for any source distribution.
\item In~\cite{Ziv90}, Ziv showed that variable-to-fixed length codes are better in a certain sense than fixed-to-variable length codes for Markov sources.  He showed this fact by using a VFSC that exploits the memory of the source. It would be a fruitful endeavor to extend Theorem \ref{thm} to the case of Markov sources to examine whether there is any difference in the first-order fundamental limits in the lossy and non-vanishing excess-distortion probability settings for fixed-to-variable and variable-to-fixed length source codes.
\end{itemize}  
%\end{remark}
\section{Achievability}\label{ach:proof}
Before proving the direct part of Theorem \ref{thm}, we first establish an intermediate result using random coding. Subsequently, in Lemmas \ref{aep} and \ref{beauty1}, we will use a randomization argument to complete the argument.
\begin{lemma}
\label{lem3e}
Under conditions (A1)--(A3), for any $N' \in \bbR^+$, there exists a sequence of  $(M',N,D,O({1}/{\ln N'}))$ variable-to-fixed source code for the number of codewords $M'$ and the code stopping time $\tau_{N'}$ satisfying
\begin{align}
\label{eq27newl}
\ln M'&= N' R(D)+O(\sqrt{N'\ln N'}),\\
\bbE(\tau_{N'})&=N'. 
\end{align} 
\end{lemma}
\begin{IEEEproof}
The proof is partly based on the standard rate-distortion proof in Cover and Thomas~\cite[Chapter 10]{Cov06} and ideas from Koga and Yamamoto~\cite{Koga2005} and Polyanskiy, Poor and Verd\'u~\cite{Yury2011}. However, we need to make some changes to the typical  set compared with the proofs in~\cite{Koga2005} and~\cite{Cov06}. It is enough to show the existence of an $(M',N',D,O({1}/{\ln N'}))$-VFSC  such that $D-(\ln N')/N'> d_{\rmm\rmi\rmn}$. To define such an online lossy source code we need to specify $(f_n,g_n)$. For convenience in the exposition in the sequel, define the  vanishing sequence
\begin{align}
\label{definegamma}
\gamma_{N'}:=\sqrt{\frac{\ln N'}{N'}}.
\end{align}
From the conditions (A2)  and  (A3), for each $N'\in\bbR^+$, we can choose a conditional distribution $P_{\hatV|V}^{(N')}(\hatv|v)$ which depends on $N'$ such that the following hold:
\begin{align}
\label{eq14cure}
I(V;\hatV)&=R(D-\gamma_{N'}),\\
\label{eq15cure}
0\leq \bbE[d(V,\hatV)]&\leq D-\gamma_{N'}.
\end{align}
Note that the expectations in~\eqref{eq14cure} and~\eqref{eq15cure} are taken with respect to the joint distribution   $P_{V\hatV}^{(N')}=P_V \times P_{\hatV|V}^{(N')}$. For each $n$, the encoding scheme is as follows. Let 
\begin{align}
\label{eq20}
P_{\hatV}^{(N')}(\hatv):=\sum_{v\in\calV}P_V(v)P_{\hatV|V}^{(N')}(\hatv|v).
\end{align}
We define $M'$ infinite-dimensional vectors $C_j = (C_{j,1}, C_{j,2}, C_{j,3},\ldots)\in\hat{\calV}^\infty$ each indexed by a $j \in \{1,\ldots, M'\}$ and which are  independently generated and distributed according to $P_{\hatV}$ on $\bbR^{\infty}$. The codebook $\{C_1,C_2,\ldots, C_{M'}\}$ is revealed to both encoder and decoder.

Define the multi-letter information density
\begin{align}
\label{informdensity}
\iota_{ P_{V\hatV}^{(N')}}(v^n,\barv^n):=\sum_{k=1}^n \iota_{ P_{V\hatV}^{(N')}}(v_k,\barv_k), \quad \forall (v^n,\barv^n) \in \calV^n \times \hat{\calV}^n\enspace \mbox{and} \enspace n\geq 1.
\end{align}
For each $(v^n,\barv^n) \in \calV^n \times\hat{\calV}^n$, define the following indicator function
\begin{align}
\label{logic}
\Psi(v^n,\barv^n):=\bone \big\{\calA(v^n,\barv^n ) \wedge\calB(v^n,\barv^n) \big\}
\end{align} 
where  the clauses
\begin{align}
\calA(v^n,\barv^n ) & := \{d(v^n,\barv^n)\leq D\}\label{eqn:defeventA} ,\quad\mbox{and}\\
\calB(v^n,\barv^n)&:= \Big\{\iota_{ P_{V\hatV}^{(N')}}(v^n;\barv^n)\leq n (R(D-\gamma_{N'})+\gamma_{N'}) \Big\}.\label{eqn:defeventB}
\end{align}
 Thus $\Psi(v^n,\barv^n)=1$ if both  $\calA(v^n,\barv^n )$ and $\calB(v^n,\barv^n )$ are true.  Also define
\begin{align}
\label{logicnew}
\tilde{\Psi}(v^n,\barv^n):=1-\Psi(v^n,\barv^n).
\end{align}
The stopping time $\tau_{N'} \in\sigma(V^n)$ is now simply defined as 
\begin{align}
\tau_{N'}:=N'. \label{eqn:tau_N}
\end{align}
Let $C_j(n):=(C_{j,1},C_{j,2},\ldots,C_{j,n})$ be a length-$n$ vector form using the infinite-dimensional vector $C_j$. The encoding function $f_n:\calV^n\to\{1,\ldots, M'\}$ for each $n\geq 1$ is defined as
\begin{align}
\label{encod}
f_n(V^n):=\left\{ \begin{array}{cc}
\max\{ j: \Psi(V^n,C_j(n))=1\} & \mbox{if } \exists\, j \mbox{ s.t. } \Psi(V^n,C_j(n))=1 \\
1 & \mbox{otherwise}
\end{array}   \right. .
\end{align} %if there exits a $j$ such that $\Psi(V^n,C_j(n))=1$, otherwise, we set $f_n(V^n)=1$. 
The decoding decision at each time $n$ is given by
\begin{align}
\label{decod}
\hatV^n=g_n(f_n(V^n))=C_{f_n(V^n)}(n).
\end{align} 
Let $V^{\infty},\barV^{\infty},\tilV^{\infty}$ be i.i.d.\ infinite dimensional vectors which are distributed according to 
\begin{align}
\label{gener}
P_{V^{\infty} \barV^{\infty} \tilV^{\infty}}(v^{\infty},\barv^{\infty},\tilv^{\infty})= \prod_{k=1}^{\infty} P_V(v_k) P_{\hatV|V}^{(N')}(\barv_k|v_k)P^{(N')}_{\hatV}(\tilv_k).
\end{align}
Now, from our encoding and decoding functions~\eqref{encod}, and~\eqref{decod} for each $n\in \bbN$, the event error is
\begin{align}
\label{goodobserv}
\{d(V^n,\hatV^n)> D\}\subset \bigcap_{j=1}^{M'}\big\{\tilde{\Psi}(V^n,C_j(n))=1\big\},
\end{align}
where~\eqref{goodobserv} follows from the choice of $j \in \{1,2,\ldots,M'\}$ as the {\em largest} index such that $d(V^n,C_j(n))\leq D$.

It follows from~\eqref{encod} and~\eqref{goodobserv} that
\begin{align}
\bbP(d(V^n,\hatV^n)> D)& \leq \bbP\bigg(\bigcap_{j=1}^{M'} \{\tilde{\Psi}(V^n,C_j(n))=1\}\bigg) \nonumber\\
&=\sum_{v^n \in \calV^n} P_{V^n}(v^n)\bbP\bigg(\bigcap_{j=1}^{M'} \{\tilde{\Psi}(v^n,C_j(n))=1\} \bigg|V^n=v^n\bigg)\\
\label{contract10}
&=\sum_{v^n \in \calV^n} P_{V^n}(v^n)\prod_{j=1}^{M'}\bbP\left( \tilde{\Psi}(v^n,C_j(n))=1 \middle|V^n=v^n\right)\\
\label{eqkey1}
&=\sum_{v^n \in \calV^n} P_{V^n}(v^n)\left[\bbP\left(\tilde{\Psi}(v^n,C_1(n))=1 \middle|V^n=v^n\right) \right]^{M'}\\
\label{eq59best}
&=\sum_{v^n \in \calV^n} P_{V^n}(v^n)\left[\bbP\big(\tilde{\Psi}(v^n,C_1(n))=1\big)\right]^{M'}\\
\label{eq62good}
&=\sum_{v^n \in \calV^n} P_{V^n}(v^n)\left[1-\bbP\big(\Psi(v^n,C_1(n))=1\big)\right]^{M'},
\end{align}
where~\eqref{contract10} follows from the fact that $C_j$'s are independent for all $j$,~\eqref{eqkey1} follows from the fact that $C_j$'s are identically distributed, and~\eqref{eq59best} follows from the fact that $V^n$ is independent of $C_j(n)$ for all $n$ and $j$.

Moreover, we also have
\begin{align}
\label{beauty1000}
&\bbP(\Psi(v^n,C_1(n))=1)  
= \bbP(\Psi(v^n,\tilV^n)=1) 
= \bbP(\Psi(v^n,\barV^n)=1)\\
%&=\bbE[\bone\{\Psi(v^n,\barV^n)=1\}]\\
&=\sum_{\barv^n \in \hat{\calV}^n} P_{\barV^n}^{(N')}(\barv^n) \bone\{\Psi(v^n,\barv^n)=1\}\\
\label{eqchange}
&=\sum_{\barv^n \in \hat{\calV}^n}P_{\barV^n|V^n}^{(N')}(\barv^n|v^n)\exp\big(-\iota_{ P_{V\hatV}^{(N')}}(v^n;\barv^n)\big)\bone\{\Psi(v^n,\barv^n)=1\}\\
&= \sum_{\barv^n\in \hat{\calV}^n}P_{\barV^n|V^n}^{(N')}(\barv^n|v^n)\exp\big(-\iota_{ P_{V\hatV}^{(N')}}(v^n;\barv^n)\big)\bone \big\{\calA(v^n,\barv^n )  \wedge \calB(v^n,\barv^n )  \big\}\\
&\geq \exp \big(-n(R(D-\gamma_{N'})+\gamma_{N'})\big)    \sum_{\barv^n\in \hat{\calV}^n}P_{\barV^n|V^n}^{(N')}(\barv^n|v^n)\bone \big\{\calA(v^n,\barv^n )  \wedge \calB(v^n,\barv^n )  \big\},
\end{align} 
where~\eqref{beauty1000} follows from the fact that $P_{\barV^n}^{(N')} = (P_{\hatV}^{(N')})^{\times n} = P_{\tilV^n}^{(N')}=P_{C_1(n)}$ (cf.~\eqref{gener}), and~\eqref{eqchange} follows from a standard change of measure argument and the definition of $\iota_{ P_{V\hatV}^{(N')}}(v^n;\barv^n)$ in \eqref{preinform} and \eqref{informdensity}. It follows that we have
\begin{align}
&\big[1-\bbP\big(\Psi(v^n,C_1(n))\big) \big]^{M'}\nonumber\\
\label{eq51new1}
&\leq 1+\exp\big[-M'\exp \big(-n(R(D-\gamma_{N'})+\gamma_{N'})\big) \big] -\sum_{\barv^n\in \hat{\calV}^n}P_{\barV^n|V^n}(\barv^n|v^n)\bone \big\{\calA(v^n,\barv^n )  \wedge \calB(v^n,\barv^n )  \big\},
\end{align}
where~\eqref{eq51new1} follows from~\cite[Lemma 10.5.3]{Cov06} (which states that for any $0\le x,y\le 1$ and $n>0$, $(1-xy)^n\le 1-x	+e^{-yn})$).

From~\eqref{eq62good},~\eqref{eq51new1}, and~\eqref{logic} we obtain that for all $n\geq 1$,
\begin{align}
&\bbP(d(V^n,\hatV^n)>D) \nn\\*
&\leq 1+\exp \big[ -M'\exp \big(-n (R(D-\gamma_{N'})+\gamma_{N'} )  \big)  \big]-\bbE[\Psi(V^n,\barV^n)]\\
&=1+\exp \big[ -M'\exp \big(-n (R(D-\gamma_{N'})+\gamma_{N'} )  \big)  \big]
\label{eqlastday}
  -\bbP \big(\calA(V^n,\barV^n)\cap\calB(V^n,\barV^n)\big)\\
\label{test}
&\leq \bbP(d(V^n,\barV^n)> D)+\bbP\Big(\big|\iota_{ P_{V\hatV}^{(N')}}(V^n;\barV^n)- nR(D-\gamma_{N'})\big|>n \gamma_{N'}\Big)\nonumber\\
&\qquad +\exp \big[ -M'\exp \big(-n (R(D-\gamma_{N'})+\gamma_{N'} )  \big)  \big],  
\end{align}
where the last inequality in~\eqref{test} holds by the union bound and the definitions of the events $\calA(V^n,\barV^n)$ and $\calB(V^n,\barV^n)$ in~\eqref{eqn:defeventA} and~\eqref{eqn:defeventB} respectively.
Now, by Chebyshev's inequality we have, for all $N'$ sufficiently large, that
\begin{align}
\label{melvin}
\bbP\Big(\big|\iota_{ P_{V\hatV}^{(N')}}(V^n;\barV^n)- nR(D-\gamma_{N'})\big|>n \gamma_{N'}\Big)&\leq \frac{\var \big(\iota_{ P_{V\hatV}^{(N')}}(V;\hatV)\big)}{n^2\gamma_{N'}^2}\\
\label{eq27}
&\leq \frac{(\rmV(D)+1) N'}{n \ln N'}, 
\end{align}
where~\eqref{eq27} follows from assumptions  (A1), (A2), and  (A3). Similarly, from Chebyshev's inequality, we also have
\begin{align}
\bbP(d(V^n,\barV^n)> D)&=\bbP \left(nd(V^n,\barV^n)-n\bbE[d(V,\hatV)]> n(D-\bbE[d(V,\hatV) ] )  \right)\\
\label{eq43}
&\leq \bbP\left(nd(V^n,\barV^n)-n\bbE[d(V,\hatV)]> n\gamma_{N'}\right)\\
&\leq \frac{\var(d(V,\hatV)) N'}{n\ln N'}\\
\label{good2}
&\leq \frac{(\tilde{\rmV}(D)+1)N' }{n \ln N'},
\end{align} where~\eqref{eq43} follows from~\eqref{eq15cure}  and~\eqref{good2} follows from~\eqref{eq11hold} and assumption~(A3).

It follows from~\eqref{definegamma},~\eqref{test},~\eqref{eq27}, and~\eqref{good2} that for all $n\geq 1$
\begin{align}
&\bbP(d(V^n,\hatV^n)>D) \leq \frac{(\tilde{\rmV}(D)+1)) N'}{n \ln N'}+\frac{(\rmV(D)+1)) N'}{n \ln N'} +\exp \big[ -M'\exp \big(-n (R(D-\gamma_{N'})+\gamma_{N'} )  \big)  \big]. \label{eqn:bound_dist}
\end{align}
By using the fact that $\tau_{N'}=N'$ (cf.\ \eqref{eqn:tau_N}) and \eqref{eqn:bound_dist}, we have 
\begin{align}
\bbP(d(V^{\tau_{N'}},\hatV^{\tau_{N'}})>D)&=\bbP(d(V^{N'},\hatV^{N'})>D)\\
\label{c1}
&\leq \frac{(\tilde{\rmV}(D)+1) N'}{n \ln N'}+\frac{(\rmV(D)+1)N'}{n\ln N'} +\exp \big[ -M'\exp \big(-n (R(D-\gamma_{N'})+\gamma_{N'} )  \big)  \big]. 
\end{align}
Now, we choose
\begin{align}
\ln M'&=N'(R(D-\gamma_{N'})+\gamma_{N'})+\ln \ln N'\\
\label{c2}
&= N' R(D)+ O(\gamma_{N'}),
\end{align} where~\eqref{c2} follows from assumption (A3) and Taylor's expansion. Combining~\eqref{c1} and~\eqref{c2}  and the fact that $\tau_{N'}=N'$  (cf.~\eqref{eqn:tau_N}) so $n=N'$, we obtain the bound
\begin{align}
\bbP(d(V^{\tau_{N'}},\hatV^{\tau_{N'}})>D) \leq O\bigg(\frac{1}{\ln N'}\bigg),
\end{align}  as $N' \to \infty$ as desired.
\end{IEEEproof}
The next two lemmas employ a randomization argument to construct a variable-to-fixed length code with the desired expected length and probability of excess distortion. In particular, we use the first $L=O(\ln  N)$ symbols of the source $\{V_n\}_{n\in\bbN}$ to generate a toss from a biased coin with success probability $p \in (\alpha,\beta]$ to follow.  See Remark~\ref{rmk:cont} for a significant simplification of the randomization argument when the source is absolutely continuous with respect to the Lebesgue measure.
\begin{lemma}\label{aep} For any memoryless source $\{V_n\}_{n\in\bbN}$ and $0<\alpha<\beta<1$,  there exists a positive integer $L$ and a subset $\calA_L$ of $\calV^{L}$ such that
\begin{align}
\int_{v^L\in\calA_L} \,\rmd P_V^{L}(v^L)&\in (\alpha, \beta],
\end{align} where the product distribution $P_V^L(v_1, \ldots,v_L):=\prod_{j=1}^L P_V(v_j)$ for all $v^L \in \calV^L$.
\end{lemma}
\begin{IEEEproof} Please refer to Appendix~\ref{app:aep}. \end{IEEEproof}
\begin{lemma} \label{beauty1}
Given a memoryless source with distribution $P_V$, under conditions (A1)--(A3), the following holds
\begin{align}
\label{ach}
R_{\rm{vf}}^*(D,\eps) \leq (1-\eps) R(D) \quad\forall\,\eps\in (0,1].
\end{align}
\end{lemma} 
\begin{IEEEproof}
Based on Lemma~\ref{lem3e}, for a sufficiently  large constant $\Gamma>0$ and average code length $N'\in\bbR^+$, we may assume the existence of  an $(M',N',D,\Gamma/\ln N') $-VFSC.  %Also fix a 
It is enough to show that for any arbitrarily small $\delta>0$, there exists an $(M,N,D,\eps)$-VFSC such that
\begin{align}
\ln  M &\leq (1-\eps)N  R(D),\\
\rvP_{\rmd}(N,D)&\leq \eps,\\
N  \leq \bbE(\tau_N) &\le  N(1+f(\delta))+1, \label{eqn:newEtau} 
\end{align} for some function $f(\delta)\to 0$ as $\delta\to 0^+$. 
We can construct such an $(M,N,D,\eps)$-VFSC with a new stopping time $\tau_N$ from the given $(M',N',D,\Gamma/\ln N') $-VFSC as follows:
\begin{itemize}
\item The source encoder choose $N\in\bbR^+$ (sufficiently large) and puts
\begin{align}
\label{eq62f}
N'&:=(1-\eps)N, \\
%a_{\eps,\delta}&:=1+\frac{\eps+\delta}{(1-\eps)\eps},\\
\label{definegamma1}
\gamma&:=1+\frac{\eps+\delta}{(1-\eps)\eps},\\
\label{definealpha}
\alpha&:=\frac{\eps}{(1-\eps)(\gamma-1)} = \frac{\eps^2}{\eps+\delta},\\
\label{70second}
\beta&:=\frac{\eps+\alpha}{2},\\
\label{69second}
L&:=\lfloor\ln N'\rfloor.
\end{align}
From these definitions, we see that $\gamma>1$, $\alpha<\beta<\eps$, $\alpha\to\eps^-$, $\beta\to\eps^-$ as $\delta\to 0^+$. 
%\begin{align}
%\gamma &> 1,\\
%\label{keyalpha}
%\alpha &<\beta,\\
%\alpha &\to \eps^{-},\enspace \mbox{as\enspace $\delta\to 0$}.
%\end{align}
\item Using Lemma \ref{aep}, we observe that the encoder can select  a subset $\calH_1:=\calA_L\subset  \calV^L$, and thus also $\calH_0:=\calV^L\setminus \calH_1$,  such that the following hold 
\begin{align}
\label{eq67good}
p&:=\int_{v^L\in \calH_1} \,\rmd P_V^L(v^L)\in (\alpha,\beta],\\
\label{eq68good}
1-p&:=\int_{v^L\in \calH_0} \,\rmd  P_V^L(v^L)=1-\int_{v^L\in \calH_1} \,\rmd P_V^L(v^L)\in (1-\beta,1-\alpha].
\end{align} 
\end{itemize}
Now, since $L=\lfloor\ln N'\rfloor<N'$ and $\gamma> 1$,  for any sequence $\{  V_n\}_{n\in\bbN}$, the encoder performs the following randomization operation:
\begin{itemize}
\item If $(V_1,V_2,\ldots, V_L )\in \calH_1$, the encoder sets $\tau_N=\lceil\gamma\tau_{N'}\rceil=\lceil\gamma N'\rceil$, indexes the sequence $V^{\tau_N}$ of length $\tau_N$ by $1 \in \{1,2,\ldots,M'\}$, and decodes it by using the given $(M',N',D,\Gamma/\ln N' )$-VFSC  for the first $N'\le\tau_N$ symbols of this sequence. This means that the decoder always assumes that each source sequence has  a {\em fixed} length $N'$. 
\item If $(V_1,V_2,\ldots,V_L)\in \calH_0$, the encoder sets $\tau_N=\tau_{N'}=N'$ and assigns the sequence $V^{\tau_N}$  an  index in $\{1,2,\ldots,M'\}$ according to the given    $(M',N',D,\Gamma/\ln N' )$-VFSC. Decoding is also done using this code.
\end{itemize}
Note that with the so-described variable-to-fixed length lossy source coding scheme, the expected  code length  satisfies
\begin{align}
\label{eq65f}
\bbE(\tau_N)&=  p \bbE(\lceil\gamma\tau_{N'}\rceil) + (1-p) \bbE(\tau_{N'})  \\
&=  p \lceil\gamma N'\rceil + (1-p) N'\\
&\in [(\gamma p+ (1-p)) N', (\gamma p+ (1-p)) N'+1] \label{eqn:pL}\\
\label{cond}
&=[(\gamma p+ (1-p)) (1-\eps)N, (\gamma p+ (1-p)) (1-\eps)N+1]\\ 
&\subset [(\gamma \alpha+ (1-\alpha))(1-\eps)N,(\gamma \beta+ (1-\beta))(1-\eps)N+1] \\
\label{nownew}
&= [N,(1+f(\delta))N+1].
\end{align} where~\eqref{cond} follows from~\eqref{eq62f},~\eqref{nownew} follows from~\eqref{definealpha} and~\eqref{70second} (note that $(\gamma\alpha + (1-\alpha))(1-\eps)=1$ and $(\gamma\beta + (1-\beta))(1-\eps)=1 + f(\delta)$), and $f(\delta)$ is some function that tends to zero  as $\delta\to 0^+$.  Hence, \eqref{gh} is satisfied (if we consider $\delta\to 0^+$). 

In addition, for $N'$ sufficiently large, the probability of excess-distortion of the proposed scheme %is upper bounded as follows:
\begin{align}
\bbP(d(V^{\tau_N}, \hatV^{\tau_N})>D)\leq \beta \cdot  1 +(1-\alpha)\cdot  \frac{\Gamma}{\ln N'} \leq \eps. \label{eqn:ep_stop}
\end{align} %where we have used~\eqref{70second} and~\eqref{69second}.

Hence, by Lemma~\ref{lem3e} and the choice of $N'$ in \eqref{eq62f}, we have
\begin{align}
\ln M&=N' R(D)+ O(\sqrt{N'\ln N'})\\
\label{goodness}
&=(1-\eps) N R(D)+ O(\sqrt{N\ln N}).
\end{align}
It follows from~\eqref{gh} and~\eqref{goodness} that~\eqref{ach} holds. This concludes the proof of Lemma~\ref{beauty1}.
\end{IEEEproof}

\begin{remark} \label{rmk:cont}
We note that if the source distribution $P_V$ is absolutely continuous with respect to the Lebesgue measure (e.g., $P_V$ is Gaussian), then the  above somewhat involved randomization argument using $V^L = (V_1,\ldots, V_L) \in\calV^L$ can be significantly simplified. In particular, we can just partition $\calV$ into two subsets $\calH_0$ and $\calH_1$ such that 
$\int_{v \in \calH_1} \rmd P_V(v )=p$; this is possible because of the continuous nature of $P_V$.  Next, when we simply look at the first symbol $V_1\in\calV$, check whether it belongs to $\calH_1$ or $\calH_0$ to decide on the stopping time ($\lceil\gamma N'\rceil$ or $N'$ in the two cases above~\eqref{eq65f}). 
\end{remark}
\section{Converse} \label{conv:proof}
\begin{lemma} \label{lemeasy} Let $\xi>0$. For any real sequence $\{x_n\}_{n\geq 1}$ such that $\lim_{n\to \infty} x_n$ exists, the following holds
\begin{align}
\label{exp}
\lim_{n\to \infty} \bone\{x_n >\xi\} \geq \bone\{\lim_{n\to \infty} x_n>\xi\}.
\end{align}
\end{lemma}
\begin{lemma} \label{conve} Given a memoryless source with distribution $P_V$, under conditions (A1)--(A2), the following holds 
\begin{align}
\label{conveeq}
R_{\rm{vf}}^*(D,\eps)  \geq (1-\eps)R(D),\quad\forall\,\eps\in (0,1].
\end{align}
\end{lemma}
\begin{IEEEproof}
For any $n\in \bbN$, we have the following Markov chain
\begin{align}
V^n-f_n(V^n)-\hatV^n. \label{eqn:mc}
\end{align}
Now, define~\cite{KostinaVL,kost17} %\red{originally not clear}
\begin{align}
\label{kost}
\rvR_{V^n}(D,\eps):=\min_{P_{\hatV^n|V^n}:\bbP(d(V^n,\hatV^n) >D) \leq \eps} I(V^n;\hatV^n).
\end{align}
It follows that for a fixed $(M,N,D,\eps)$-VFSC, and for some $B_{\eps}>0$, the following inequalities hold  for all $n\geq 1$:
\begin{align}
\ln M &\geq H(f_n(V^n))\\
&=I(f_n(V^n);V^n)\\
&\geq I(V^n;\hatV^n) \label{eqn:use_mc}\\
\label{gstr}
&\geq \max\{0,\rvR_{V^n}(D,\eps)\bone\{\bbP(d(V^n,\hatV^n)> D)\leq \eps\}\}\\
\label{victor}
&=\max\{0,\rvR_{V^n}(D,\eps)\bone\{\bbP(d(V^n,\hatV^n)\leq D)>1-\eps\}\}\\
\label{best1}
&\geq \max\Big\{0, \big[(1-\eps)nR(D)-B_{\eps}\sqrt{n} \big] \bone\{\bbP(d(V^n,\hatV^n)\leq D)>1-\eps\}\Big\}.
\end{align} Here, \eqref{eqn:use_mc} follows from~\eqref{eqn:mc},~\eqref{gstr} follows from~\eqref{kost}  and~\eqref{best1} holds for all $n$ sufficiently large due to the main results in Kostina {\em et al.}~\cite{KostinaVL,kost17}. Note that under our assumptions (A1) and (A2), the conditions ($\rmA_1$)--($\rmA_5$)  in~\cite{kost17} are satisfied.

Let $a\wedge b := \min\{a,b\}$. It follows from~\eqref{best1} that
\begin{align}
\label{best2}
&\ln M \geq \max \left\{0, \big[(1-\eps)(n\wedge \tau_N) R(D)-B_{\eps}\sqrt{n\wedge \tau_N}\big]\bone \big\{\bbP(d(V^{n\wedge \tau_N},\hatV^{n\wedge \tau_N})\leq D)>1- \eps \big\}\right\}.
\end{align}
From~\eqref{gh}, we have that $\bbP(\tau_N   <\infty)=1$. Hence we obtain that 
\begin{align}
\lim_{n\to \infty} (1-\eps)(n\wedge \tau_N) R(D)-B_{\eps}\sqrt{n\wedge \tau_N}=(1-\eps)\tau_N R(D) -B_{\eps}\sqrt{\tau_N},  \quad\mbox{a.s.}
\label{goodtrick}
\end{align}
and 
\begin{align}\label{good0}
\lim_{n\to \infty} \bone\{\bbP(d(V^{n\wedge \tau_N},\hatV^{n\wedge \tau_N})>D)\leq \eps\}&\geq \bone\{\lim_{n\to \infty} \bbP(d(V^{n\wedge \tau_N},\hatV^{n\wedge \tau_N})\leq D)>1- \eps\}\\
\label{good1}
&=\bone\{\bbP(d(V^{\tau_N},\hatV^{\tau_N})\geq D)>1- \eps\}\\
&=\bone\{\bbP(d(V^{\tau_N},\hatV^{\tau_N})> D)\leq \eps\}\\
\label{best3}
&=1\quad\mbox{a.s.}
\end{align}
In the above chain,~\eqref{good0} follows from Lemma~\ref{lemeasy} and~\eqref{good1} follows from the fact that
\begin{align}
\label{hold}
\lim_{n\to \infty} \bbP(d(V^{n\wedge \tau_N},\hatV^{n\wedge \tau_N})>D)=\bbP(d(V^{\tau_N},\hatV^{\tau_N})>D)
\end{align} from an application of the dominated convergence theorem~\cite{Royden} and the fact that $\bbP(\tau_N<\infty)=1$. Finally,   \eqref{best3} follows from~\eqref{errordef}.  

By taking limit  as  $n\to\infty$ on the right-hand-side of~\eqref{best2} ($\ln M$ is not a function of $n$) and using~\eqref{best3}, we obtain that the following holds almost surely   
\begin{align}
\ln M \geq \max \big\{0,(1-\eps)\tau_NR(D)  -B_{\eps}\sqrt{\tau_N}\big\}.
\end{align}
Taking expectations  on both sides, it follows that
\begin{align}
\ln M &\geq \max\big\{0, \bbE[(1-\eps)\tau_NR(D) -B_{\eps}\sqrt{\tau_N}]\big\}\\
&=\max\big\{0,(1-\eps)\bbE[\tau_N]R(D)-B_{\eps}\bbE[\sqrt{\tau_N}]\big\}\\
\label{test1}
&\geq \max\big\{0,(1-\eps) \bbE[\tau_N]R(D)-B_{\eps} \sqrt{\bbE(\tau_N)}\big\}\\
\label{test2}
&= (1-\eps)NR(D)-O(\sqrt{N}).
\end{align}
Here,~\eqref{test1} follows from the concavity of the function $g(x):=\sqrt{x}$ and~\eqref{test2} follows from~\eqref{gh} and the fact that the function $\tilg(x):=(1-\eps)x R(D)-B_{\eps}\sqrt{x}$ is non-decreasing for $x$ large enough. It follows from~\eqref{test2} and~\eqref{gh} that~\eqref{conveeq} holds. This concludes the proof of Lemma~\ref{conve}.
\end{IEEEproof}
\appendices
\section{Proof of Lemma \ref{aep}} \label{app:aep} 
%\begin{IEEEproof}[Proof of Lemma~\ref{aep}] 
From the asymptotic equipartition property (AEP)~\cite[Theorem 3.1.2]{Cov06}, we can find a  sufficiently large positive integer $L$ and a subset $\calT_L \subset\calV^L$ such that 
\begin{align}
\label{newgood1}
\bbP(V^L \in \calT_L)>\alpha.
\end{align}
By Zorn's Lemma~\cite{Royden} (partially ordered sets by inclusion), we can find a subset $\calA_L$ of $\calV^L$ which has the smallest size among all subsets $\calT_L$ of $\calV^L$ which satisfy~\eqref{newgood1}.
It is enough to show that
\begin{align}
\label{star}
\bbP(V^L \in \calA_L)\leq \beta.
\end{align}
Now, we will show that~\eqref{star} holds by contradiction. Indeed, assume to the contrary, that $\bbP(V^L \in \calA_L)>\beta$. Then, choose a subset $\mathcal{\tilA}_L$ of $\calA_L$ which has the number of elements
\begin{align}
\label{goodfact}
|\mathcal{\tilA}_L|=\bigg\lceil\frac{\alpha}{\beta}|\calA_L|\bigg\rceil,
\end{align} %and that the subset $\mathcal{\tilA}_L$ 
also the property that $\mathcal{\tilA}_L$  has largest probability among all the subsets of $\calA_L$ of the same size given by \eqref{goodfact}. It follows that
\begin{align}
\bbP(V^L \in \mathcal{\tilA}_L)&\geq \frac{\big\lceil\frac{\alpha}{\beta}|\calA_L|\big\rceil}{|\calA_L|} \bbP(V^L \in \calA_L)
> \frac{\alpha}{\beta}\cdot\beta
=\alpha.
\end{align}
But, we know from~\eqref{goodfact}  and~\cite[Theorem 3.3.1]{Cov06} which states that $|\calA_L|$ (satisfying \eqref{newgood1}) is of exponential size that
\begin{align}
\label{contra}
|\mathcal{\tilA}_L| <\frac{\alpha}{\beta}|\calA_L|+1<|\calA_L|
\end{align} for $L$ sufficiently large. %, where~\eqref{contra} follows from~\cite[Theorem 3.3.1]{Cov06} which states that $|\calA_L|$ has exponentially number of elements. 
Inequality~\eqref{contra} is a contradiction to our assumption about the choice of the set $\calA_L$, i.e., that it is the {\em smallest} set satisfying~\eqref{newgood1}. This concludes the proof of Lemma~\ref{aep}.
%\end{IEEEproof}

\section{Proof of Lemma  \ref{lemeasy} } \label{app:prf} 
Assume that 
\begin{align}
\label{limext}
\lim_{n\to \infty} x_n =\zeta.
\end{align}
For the case $\zeta\le\xi$,~\eqref{exp} trivially holds. Now, consider the case $\zeta>\xi$. Then, there exists an $\epsilon>0$ such that
\begin{align}
\label{easy1}
\zeta-\epsilon>\xi.
\end{align}
On the other hand, from~\eqref{limext} there exists an $N_{\epsilon} \in \bbN$ such that
\begin{align}
\label{easy2}
x_n \geq \zeta-\epsilon,
\end{align} for all $n\geq N_{\epsilon}$. From~\eqref{easy1} and~\eqref{easy2} we obtain
\begin{align}
\label{easy3}
x_n >\xi 
\end{align} for all $n\geq N_{\epsilon}$. It follows from~\eqref{easy3} that for $\zeta>\xi$, we have
\begin{align}
\lim_{n\to \infty} \bone\{x_n >\xi\} =1 =\bone\{\lim_{n\to \infty} x_n>\xi\}.
\end{align}
This concludes the proof of Lemma~\ref{lemeasy}.

\subsection*{Acknowledgements}
The authors thank Prof.\ Victoria Kostina for pointing us to several useful references.
\bibliographystyle{unsrt}
\bibliography{isitbib}
\end{document}